# Towards robust radiomics and radiogenomics predictive models for brain tumor characterization

**Maria Nadeem[1,#], Asma Shaheen, PhD[2,#], Muhammad F.A. Chaudhary, PhD[3], and Hassan Mohy-ud-Din, PhD[4,#,*]**

[1] Department of Mathematics, School of Science and Engineering, LUMS, Lahore, 54792, PAK
[2] Department of Mathematics, Computer Science, and Physics, University of Udine, Udine, 33100, Italy
[3] The Roy J. Carver Department of Biomedical Engineering, The University of Iowa, Iowa City, IA, 52242, USA
[4] Department of Electrical Engineering, School of Science and Engineering, LUMS, Lahore, 54792, PAK

[#] denote equal contributions
[*] denote corresponding author

E-mail: hassan.mohyuddin@lums.edu.pk

**Abstract.** Several (mini)processes along a medical imaging pipeline induce variability in radiomics features which ultimately compromises predictive performance on downstream tasks. In the context of brain tumor characterization, we focused on two key questions which, to the best of our knowledge, have not been explored so far: (a) stability of radiomics features to variability in multiregional segmentation masks obtained with fully-automatic deep segmentation methods and (b) subsequent impact on predictive performance on downstream tasks – such as IDH prediction and Overall Survival (OS) classification. We further constrained our study to limited computational resources setting which are found in underprivileged, remote, and/or resource-starved clinical sites in developing countries. We employed seven state-of-the-art CNNs which can be trained with limited computational resources (GPU with 12 GB RAM) and have demonstrated superior segmentation performance on Brain Tumor Segmentation challenge (BraTS). Stability of radiomics features across fully-automatic deep segmentation networks was quantified with OCCC. Subsequent selection of discriminatory features was done with RFE-SVM and MRMR feature selection methods. Predictive performance of radiomics and radiogenomics models was quantified with AUC. Our study revealed that highly stable radiomics features were: (1) predominantly texture features (79.1%), (2) mainly extracted from WT region (96.1%), and (3) largely representing T1Gd (35.9%) and T1 (28%) sequences. The average stability, in terms of OCCC, for each feature category was as follows: $0.87 \pm 0.12$ for WT, $0.76 \pm 0.13$ for TC, $0.72 \pm 0.13$ for ENC, and $0.72 \pm 0.11$ for Shape. Shape features and radiomics features extracted from the ENC tumor subregion had the lowest average stability. Stability filtering minimized non-physiological variability in predictive models as indicated by an order-of-magnitude decrease in the relative standard deviation (RSD) of AUCs. The non-physiological variability is attributed to variability in multiregional segmentation maps obtained with fully-automatic CNNs. Stability filtering significantly improved predictive performance on the two downstream tasks, i.e., IDH prediction and Overall Survival classification, substantiating the inevitability of learning novel radiomics and radiogenomics models with stable discriminatory features. The study (implicitly) demonstrates the importance of suboptimal deep segmentation networks which can be exploited as auxiliary networks for subsequent identification of radiomics features stable to variability in automatically generated multiregional segmentation maps.

# 1. Introduction

Gliomas are lesions which originate in the glial cells of the brain and account for 30% of all brain and central nervous system tumors, and 80% of all malignant brain tumors (Goodenberger and Jenkins, 2012). Gliomas can be classified into four grades, i.e. Grade I − IV, based on tumor progression indicated by the mitotic activity, cell morphology, and molecular markers (Amin et al., 2017). Tumor grade is a measure of tumor tissue abnormality compared to normal tissue. Grade I gliomas are benign and curative while Grade II and III gliomas, also known as Low-Grade Gliomas (LGGs), are relatively slow growing, having a life expectancy of several years, but may recur after treatment (Amin et al., 2017). Lastly, Grade IV gliomas, also called High-Grade Gliomas (HGGs), are of the most aggressive form and highly malignant, with a median survival of ~15 months, despite medication and treatment (Bi and Beroukhim, 2014). Unfortunately, 70% of LGGs have the tendency to grow and transform into HGGs (Maher et al., 2001).

Based on molecular markers, the World Health Organization (WHO) further categorized LGGs into various histological subtypes such as astrocytomas (AST), oligodendrogliomas (OLG), or mixed gliomas/oligoastrocytomas (OAST) (Louis et al., 2016; Rudie et al., 2019). Studies have shown that there are multiple gene mutations which, in combination, can fuel the progression of a brain tumor (Barnholtz-Sloan et al., 2008; Brennan et al., 2013). The type of mutation present in tumor cells may vary, even among tumors of the same grade e.g. AST and OLG are both Grade II gliomas but demonstrate different type of gene mutation; see Figure 1 in (Louis et al., 2016) and Figure 3 in (Rudie et al., 2019). Moreover, it is observed that in LGGs, IDH (isocitrate dehydrogenase) protein gets mutated more commonly than in HGGs. Research reveals that only 12% of Glioblastomas (GBM, Grade IV) and $50-80$% of LGGs have IDH mutation (Parsons et al., 2008; Eckel-Passow et al., 2015). WHO has recognized IDH mutation along with 1p19q chromosome arm co-deletion as a key prognostic molecular marker for the diagnosis of OLG (Grade II). On the other hand, tumors with non-mutated IDH protein, also known as IDH-wild type tumors, and 1p19q intact chromosome arm tend to have poor prognosis as compared to IDH-mutant (Cohen et al., 2013; Ceccarelli et al., 2016) mostly because the lack of mutation makes the therapeutic treatments less effective (Songtao et al., 2012; Weller et al., 2013).

3D Multi-parametric Magnetic Resonance Imaging (3D mpMRI) is commonly used to noninvasively identify and detect brain gliomas. Unlike X-ray and CT imaging, MRI provides high resolution 3D images without using ionizing radiation. Moreover, a variety of MRI sequences can be used to generate multiparametric 3D images without incurring patient dose. These multi-parametric scans provide useful complementary information for accurate characterization of tumor subregions (Menze et al., 2014). 3D mpMRIs include T1-weighted, T1-weighted contrast-enhanced (T1-Gd), T2 weighted, and Fluid attenuated inversion recovery (FLAIR) sequences. FLAIR and T2 scans highlight the whole tumor and tumor core regions respectively. T1 and T1-Gd scans highlight the enhancing tumor region and cystic/necrotic components of tumor core. The gold standard approach of segmenting tumor subregions on 3D mpMRI scans is manual segmentation performed by expert neuroradiologists. Manual segmentation is time intensive and suffers from inter-observer variability. Recently, deep learning algorithms have emerged as a promising solution for automatic segmentation of tumor subregions on multi-parametric MRI scans (Kamnitsas et al., 2016, 2017; Pereira et al., 2016; Dong et al., 2017; Havaei et al., 2017; Wang et al., 2018; Isensee et al., 2019; Myronenko, 2019; Luo et al., 2021; Bukhari and Mohy-ud-Din, 2021).

Following the 2016 WHO guidelines (Louis et al., 2016), diagnosis of brain tumors involves biopsy-driven histopathology followed by molecular analysis. A routine clinical workflow for the diagnosis of brain gliomas can be briefly described as follows (Forst et al., 2014; Louis et al., 2016): for an incoming subject, 3D mpMRI scans are acquired for a noninvasive visualization and spatial localization of brain tumor. These scans are also used to segment tumor subregions either manually, by trained neuroradiologists, or automatically, via deep learning algorithms. A small sample of the abnormal tissue is surgically removed (i.e. biopsied) followed by a detailed microscopic examination which is summarized in a pathology report. For instance, with oligodendrogliomas, histopathological findings will include, amongst others, a honeycomb like arrangement of evenly spaced tumor cells with uniformly rounded nuclei and clear haloes and capillaries with a chicken-wire appearance (Forst et al., 2014).For diffuse astrocytomas, histopathological findings will include, amongst others, well-differentiated embedded neoplastic fibrillary astrocytes with moderately increased cellularity, mild nuclear atypia having elongated nuclei, absence of necrosis, and microvascular proliferation (Lind-Landström et al., 2012; Forst et al., 2014). This is followed by extracting molecular markers which include, amongst others, IDH mutation, 1p19q chromosome arm co-deletion, ATRX mutation, TP53 mutation etc. Molecular markers provide improved diagnostic and prognostic accuracy and treatment response assessment (Louis et al., 2016; Rudie et al., 2019).

The routine clinical approach makes biopsy a must, which is not always feasible as some tumors cannot be surgically resected due to high penetration. Biopsy-guided histopathology and molecular analysis is also costly and not available in every hospital. Studies have also reported that up to 15% of IDHmutant tumors are not detectable via the standard IDH1 antibody test (Cryan et al., 2014; Gutman et al., 2015). According to a recent research article, (Beiko et al., 2014), subtotal or complete surgical removal has a huge impact on the patient's overall survival rate. IDH-mutated gliomas are more amenable to complete resection, improving patient survival, unlike IDH-wild type which show no additional benefit with further resection of the non-enhancing component of the tumor (Beiko et al., 2014). It also suggests that when dealing with malignant AST based on IDH1 status, individualized surgical strategies can be adopted which makes it even more crucial to preoperatively specify the IDH status.

Limitations of the prevalent clinical approach are significantly pronounced in underprivileged, remote, and/or resource-starved clinical sites in developing countries. More specifically, in developing countries, there are very few clinical sites equipped with histopathology and molecular analysis, expert neurooncologists and neuroradiologists. Tissue samples are transported to distant clinical sites, equipped with state-of-the-art histopathology and molecular facilities which consume a significant amount of time. Moreover, the total cost of a diagnostic exam is quite high; well above the average monthly income of a middle-class individual feeding, on average, six family members. This presses the need of alternative approaches for noninvasive and economical brain tumor characterization in a timely manner.

The seminal work of (Aerts et al., 2014) ushered in the era of radiomics (and radiogenomics) in computational clinical and translational imaging (Lambin et al., 2012; Mazurowski, 2015; Gillies et al., 2016; Bera et al., 2022). Radiomics is a computational imaging approach that makes non-invasive prediction from (radiological) imaging scans only. Predictions can be qualitative or quantitative. e.g., tumor present vs tumor absent, overall survival, etc. Likewise, radiogenomics is a computational imaging approach that makes non-invasive prediction of tumor genotype from (radiological) imaging scans only. e.g., IDH mutation, 1p19q co-deletion, etc. A commonly followed radiogenomics workflow is shown in Figure 1. Firstly, 3D mpMRI scans are acquired, on which tumor segmentation is performed, either manually by experienced neuroradiologists or by using automatic deep learning algorithms, to identify tumor subregions including edema, necrosis, enhancement, and non-enhancement area. Multi-region segmentation maps are used to mine a large set of imaging features including shape, intensity, texture, histogram, and volumetric features. A combination of feature selection techniques is used to select an optimal subset of features (of a pre-specified cardinality) with high predictive power (Rakotomamonjy, 2003; Kursa and Rudnicki, 2011; Venkatesh and Anuradha, 2019; Yuan et al., 2019; Solorio-Fernández et al., 2020). The optimal subset of features is used to train a classifier that yields a robust and accurate radiomics (or radiogenomics) signature (Zhang et al., 2017; Bae et al., 2018; Li et al., 2018; Lu et al., 2018; Wu et al., 2019)

Mining stable radiomics features has been a major impediment towards routine adoption of computational solutions in hospitals and clinical sites. Several steps (or mini-processes) along a medical imaging pipeline induce strong variability in radiomic features rendering the learned radiomics (or radiogenomics) signature unreliable. Variability-inducing modules include image acquisition, scanner, image reconstruction protocol, post-processing, segmentation, and feature extraction software. As of now, variability of radiomic features across feature extraction softwares (i.e., unique in-house implementations) is resolved with the Image Biomarker Standardization Initiative (IBSI) (Zwanenburg et al., 2020). IBSI standardized more than hundred radiomics features with excellent reproducibility across multiple imaging modalities.

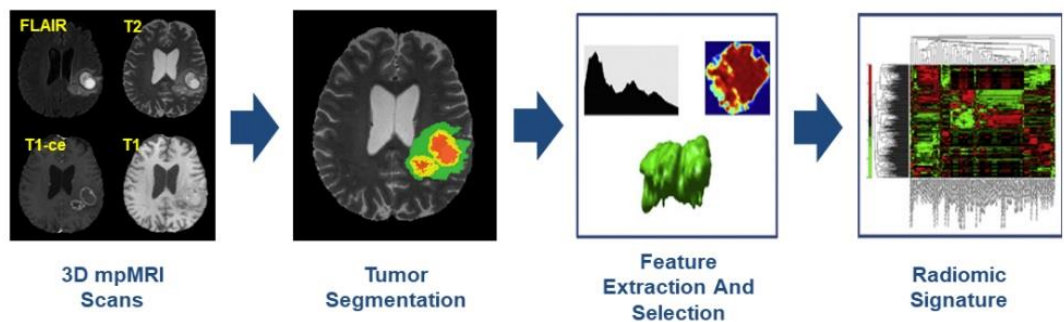

**Figure 1.** *Computational approach for radio(geno)mics of brain gliomas*: After acquiring 3D mpMRI scans, segmentation of tumor subregions is performed either manually or using (semi) automatic algorithms. The (segmented) tumor subregion masks are used to extract region-specific image-guided features (such as intensity, texture, shape, etc.) from 3D mpMRI scans. A small subset of features with high predictive power is selected from this collection using feature selection approaches. Finally, a predictive model, i.e., a radio(geno)mics signature, is learned which associates the tumor phenotypes to a specific genotype. This figure is inspired by (Lambin et al., 2012).

Several studies have explored stability of radiomic features to variations in image acquisition (voxel resolution, slice thickness, image contrast) (Zhao et al., 2016; Berenguer et al., 2018; Suter et al., 2020; Verma et al., 2022), scanners (magnetic field strength, echo time and repetition time) (Styner et al., 2002; Stonnington et al., 2008; Saha et al., 2017; Um et al., 2019; Suter et al., 2020; Verma et al., 2022), reconstruction protocols (iterative and non-iterative algorithms) (Kim et al., 2016; Zhao et al., 2016; Choe et al., 2019; Meyer et al., 2019), post-processing (noise reduction, intensity normalization, bias field correction, co-registration, skull stripping) (Shiri et al., 2020), and segmentation algorithms (manual contouring, semi-automatic contouring, and probabilistic segmentation) (Lee et al., 2017; Tixier et al., 2019; Haarburger et al., 2020). Most of these studies have only focused on the stability and reproducibility of radiomic features and not on the stability of radiomics (and radiogenomics) predictions on downstream tasks (Kalpathy-Cramer et al., 2016a; Haarburger et al., 2020).

An important exception is an extensive empirical study of (Suter et al., 2020) which quantified stability of radiomics features across 125 distinct perturbations (image acquisition, reconstruction, and post-processing) followed by analyzing the stability of radiomics prediction for overall survival classification task. The stability of radiomics features was quantified with Intraclass Correlation Coefficient (ICC). ICC assumes values between zero and one (inclusive) with ICC = 0 implying unstable feature and ICC = 1 implying (perfectly) stable feature. (Suter et al., 2020) made two important conclusions (Shaheen et al., 2022): Shape features were most robust (ICC ∈ [0.97, 0.99]), followed by first order features (ICC ∈ [0.48, 0.92]), texture features (ICC = [0.28, 0.83]), and deep features (ICC ∈ [0.48, 0.86]). Moreover, radiomics signature learned with (stable) shape features yielded superior prediction performance on overall survival classification task. (Suter et al., 2020) used BraTumIA software (Porz et al., 2014) for fully automatic segmentation of tumor subregions on 3D mpMRI scans and, hence, did not study stability of features to variability in automatic segmentation. (Suter et al., 2020) is also probably the first study which attributes poor predictive performance of radiomics signatures (learned with deep features) to limited stability of deep features (Shaheen et al., 2022).

As described in Figure 2, segmentation of tumor subregions on 3D mpMRI scans is a penultimate step towards mining radiomics features, which are subsequently used to learn a radiomics (or radiogenomics) signature. Robustness to variability in segmentation for brain tumor characterization has been explored by (Lee et al., 2017; Tixier et al., 2019). (Tixier et al., 2019) studied stability of radiomics features to variations in segmentation by employing two methods for segmentation of Whole Tumor region on FLAIR scans only. The first method is a semi-automatic segmentation approach where a user identifies tumor region (foreground) and background region, on FLAIR scans, with brush strokes followed by training an SVM classifier. The second method is an interactive segmentation approach where a user is recursively prompted to mark difficult-to-segment regions as either tumor or background followed by morphological post-processing and manual corrections. The authors concluded that GLCM and texture features were most robust to variations in segmentation across two segmentation approaches. Moreover, compared to semi-automatic segmentation method, interactive segmentation approach subsequently yielded more robust radiomics features. However, the authors did not study the stability of multi-regional radiomics features obtained with multi-region segmentation maps from fully automatic deep learning algorithms. (Lee et al., 2017) studied stability of radiomics features to variations in segmentation by employing two semi-automatic segmentation methods for segmentation of necrotic and enhancing region on T1-Gd scans and non-enhancing region on FLAIR/T2 scans. In the first method, a user identified tumor region (foreground) and background by drawing sketches, using 3D Slicer, followed by applying Grow-Cut algorithm. In the second method, a user delineated an initial ROI within the tumor using TumorPrism3D followed by automatic tumor segmentation using deformable model-based algorithm. The conclusion was that first order features were highly stable across the two semi-automatic segmentations methods.

In this manuscript, we conducted an extensive empirical study to quantify stability of radiomics features against variability in segmentations obtained with state-of-the-art fully automatic deep segmentation networks which are exclusively designed for multi-class brain tumor segmentation. We further explore the stability of downstream prediction models including radiogenomics signature for IDH prediction task and radiomics signature for overall survival (OS) classification task. This paper is organized as follows: Section 2 describes the empirical framework including data cohorts for segmentation, radiomics, and radiogenomics prediction tasks, fully automatic deep segmentation models, radiomics features extraction with Pyradiomics, reliability analysis, training and inference pipelines, and quantitative and statistical evaluation metrics. Section 3 presents the results of our extensive empirical study followed by an elaborate discussion in Section 4. Section 5 summarizes the conclusions of our study.

## 2. Methods

*2.1. Data cohort*

We used publicly available datasets for learning segmentation and radio(geno)mics predictive models. The datasets are summarized below:

*2.1.1. Dataset for Segmentation.* For segmentation, we used the publicly available BraTS 2020 dataset[a] (Clark et al., 2013; Menze et al., 2014; Bakas et al., 2017) comprising of a training cohort and a validation cohort. The training cohort consisted of 369 subjects with preoperative multiparametric 3D MRI scans (T1, T1-Gd, T2, and FLAIR sequences) and manual segmentation of tumor subregions (including peritumoral edema, non-enhancing core, and enhancing core) by expert neuroradiologists (Menze et al., 2014). Out of 369 subjects, 293 are HGGs and 76 are LGGs. The validation cohort consisted of 125 subjects and, unlike the training cohort, only included preoperative multiparametric 3D MRI scans. The BraTS 2020 dataset did not provide clinical and genomic information for subjects in the training and validation cohorts.

*2.1.2. Dataset for IDH prediction.* The BraTS 2020 dataset included subjects from The Cancer Imaging Archive (TCIA) (Clark et al., 2013; Menze et al., 2014; Bakas et al., 2017) and provided a name mapping file that matches the BraTS 2020 subject IDs with the TCIA subject IDs. Clinical and genomic information of subjects in TCIA can be obtained from The Cancer Genome Atlas (TCGA) (Clark et al., 2013). With the help of matched TCIA subject IDs, we managed to extract clinical and genomic information, including age, Kranofsky Performance Score (KPS), gender, tumor grade, and IDH mutation status for 148 subjects in the training cohort and 70 subjects in the validation cohort. Collectively, these 218 subjects formed the analysis cohort for our IDH prediction study; where 148 subjects were used for discovery and 70 subjects were used for testing.

The discovery dataset of 148 subjects (64 LGGs, 83 HGGs, 1 NA) included 57 subjects with IDH mutations (IDHmutant) and 91 subjects with IDH wild type (IDHwt). Manual segmentation of tumor subregions (peritumoral edema, non-enhancing core, and enhancing core) for the discovery dataset was obtained from the BraTS 2020 dataset (Clark et al., 2013; Menze et al., 2014). The testing dataset of 70 subjects (42 LGGs, 28 HGGs) include 32 subjects with IDH mutations (IDHmutant) and 38 subjects with IDH wild type (IDHwt). Segmentation of tumor subregions for the testing dataset was generated using fully automatic deep segmentation methods (see Section 2.2).

**Table 1a.** Characteristics of the dataset in the IDH classification study.

| Characteristics | Discovery Dataset | Testing Dataset | Total | $p$-values |
|---|---|---|---|---|
| **Subjects** | 148 | 70 | 218 | |
| **Tumor Grade** | 147 | 70 | 217 | 0.06 |
| Grade II | 27 | 20 | 47 | |
| Grade III | 37 | 22 | 59 | |
| Grade IV | 83 | 28 | 111 | |
| **Histology** | 147 | 70 | 217 | 0.09 |
| Astrocytoma | 21 | 11 | 32 | |
| Oligoastrocytoma | 18 | 10 | 28 | |
| Oligodendroglioma | 25 | 21 | 46 | |
| Glioblastoma | 83 | 28 | 111 | |
| **Age (years)** | 147 | 70 | 217 | 0.20 |
| Range | 18 − 84 | 23 − 80 | 18 − 84 | |
| Mean ± SD | 51 ± 16 | 54 ± 13.7 | 52 ± 15 | |
| **KPS** | 115 | 42 | 157 | 0.04 |
| Range | 50 − 100 | 40 − 100 | 40 − 100 | |
| Mean ± SD | 83.9 ± 12 | 77.8 ± 17.2 | 82.3 ± 13.8 | |
| **Gender** | 147 | 70 | 217 | 0.54 |
| Male | 76 | 40 | 116 | |
| Female | 71 | 30 | 101 | |
| **IDH status** | 148 | 70 | 218 | 0.38 |
| IDH-wild type | 91 | 38 | 129 | |
| IDH-mutant | 57 | 32 | 89 | |
| NOTES: $p < 0.05$ is considered statistically significant. | | | | |

---

[a] https://www.med.upenn.edu/cbica/brats2020/data.html

*2.1.3. Dataset for Overall survival prediction.* For Overall Survival (OS) classification task, we focused only on subjects with HGGs. Out of 293 subjects in the BraTS 2020 training cohort, complete survival information was provided for 236 subjects and Gross Tumor Resection (GTR) status was provided only for 118 subjects. Of the 118 subjects, 42 are short-term survivors, 30 are medium-term survivors, and 46 are long-term survivors. With the help of matched TCIA subject IDs, we managed to extract survival information and clinical variables of 31 HGGs from the validation cohort of the BraTS 2020 dataset. Of the 31 subjects, 16 are short-term survivors, 3 are medium-term survivors, and 12 are long-term survivors. These 149 subjects form the data cohort for our OS study where 118 subjects are used as a discovery dataset and 31 subjects are used as a testing dataset.

**Table 1b.** Characteristics of the dataset in the Overall Survival classification study.

| Characteristics | Discovery Dataset | Testing Dataset | Total | $p$-values |
|---|---|---|---|---|
| **Subjects** | 118 | 31 | 149 | |
| **Age (years)** | 118 | 31 | 149 | 0.25 |
| Range | $27.8 - 86.6$ | $17.0 - 80.0$ | $17.0 - 86.6$ | |
| Mean ± SD | $61.9 \pm 12$ | $58.4 \pm 15.5$ | $61.2 \pm 12.9$ | |
| Median | 63.5 | 58 | 62.2 | |
| **Survival groups** | 118 | 31 | 149 | 0.40 |
| Range (days) | $12 - 1767$ | $16 - 1215$ | $12 - 1767$ | |
| Mean ± SD (days) | $446.4 \pm 343.8$ | $390.8 \pm 314.4$ | $434.8 \pm 338.7$ | |
| Median (days) | 374.5 | 293.7 | 359 | |
| Short-term [< 10 months] | 42 | 16 | 58 | |
| Medium-term [10 − 15 months] | 30 | 3 | 33 | |
| Long-term [ > 15 months] | 46 | 12 | 58 | |
| NOTES: $p < 0.05$ is considered statistically significant. | | | | |

In summary, the data cohort used in this study met the following criteria: (a) histopathology confirmed primary grade II − IV glioma according to WHO criteria (Louis et al., 2016), (b) molecularly established IDH mutation (or OS) status, and (c) availability of four 3D MRI sequences including T1-weighted, T1-weighted contrast enhanced (T1-Gd), T2-weighted (T2), and Fluid Attenuated Inversion Recovery (FLAIR) images. A retrospective study on the publicly available BraTS 2020 and TCIA/TCGA datasets did not require IRB approval. Table 1a and Table 1b provides a detailed summary of the demographic, clinical, and genomic characteristics of the discovery and testing datasets used in our study.

**Table 2a.** Distribution of the dataset in the IDH classification study across diverse institutions.

| Clinical Center | Discovery Dataset (# subjects) | Testing Dataset (# subjects) | Total Subjects |
|---|---|---|---|
| MD Anderson Cancer Center | 19 | 3 | 22 |
| Henry Ford Hospital | 61 | 28 | 89 |
| UCSF | 8 | 6 | 14 |
| Duke | 5 | 2 | 7 |
| Emory University | 4 | 2 | 6 |
| Case Western | 13 | 5 | 18 |
| Case Western – St Joes | 13 | 16 | 29 |
| FINCB | − | 3 | 3 |
| Thomas Jefferson University | 25 | 5 | 30 |
| NOTES: UCSF, University of California, San Francisco | | | |
| FINCB, Fondazione IRCCS Instituto Neuroligico C. Besta | | | |

*2.1.4. Preprocessing.* The analysis cohort used in this study was acquired at nine different clinical centers across the United States with diverse imaging protocols and vendors (see Table 2a and Table 2b). Since the 3D mpMRI scans are acquired with different scanners and acquisition protocols, data preprocessing becomes imperative towards learning a physiologically meaningful radio(geno)mics signature (Madabhushi and Udupa, 2005; Um et al., 2019; Beig et al., 2020). Preprocessing of 3D MRI scans for each subject included, skull-stripping, affine registration to the SRI24 template, resampling to an isotropic $1 \times 1 \times 1$ mm$^3$ resolution, N3 bias correction, and mean-variance normalization (Sled et al., 1998; Bakas et al., 2017; Carré et al., 2020). N3 bias correction was not performed for tumor segmentation task as recommended by numerous CNN-based segmentation algorithms (Akkus et al., 2017b; Bakas et al., 2017; Isensee et al., 2019; Bukhari and Mohy-ud-Din, 2021).

**Table 2b.** Distribution of the dataset in the Overall Survival classification study across diverse institutions.

| Clinical Center | Discovery Dataset (# subjects) | Testing Dataset (# subjects) | Total Subjects |
|---|---|---|---|
| CBICA UPenn | 94 | — | 94 |
| MD Anderson Cancer Center | — | 3 | 3 |
| Henry Ford Hospital | 1 | 11 | 12 |
| UCSF | 7 | 10 | 17 |
| Duke | 7 | 2 | 9 |
| Emory University | 2 | 1 | 3 |
| Case Western | — | 1 | 1 |
| FINCB | — | 3 | 3 |
| Others (NA, MDA, UAB, WashU) | 7 | — | 7 |
| NOTES: UCSF, University of California, San Francisco<br>FINCB, Fondazione IRCCS Instituto Neuroligico C. Besta | | | |

*2.2. Tumor Segmentation*

In brain gliomas, the 3D brain tumor volume can be partitioned into non-overlapping or overlapping tumor subregions. A non-overlapping partition includes three physiological regions namely, enhancing core (ENC), peritumoral edema (PTE), and non-enhancing core (NEC) (Bakas et al., 2017). The three non-overlapping tumor subregions can be combined to yield an overlapping partition of brain tumor volume as follows (Menze et al., 2014): (a) whole tumor (WT) including all non-overlapping subregions (ENC, PTE, and NEC), (b) tumor core (TC) including ENC and NEC, and (c) enhancing tumor (ENC). The overlapping partition (i.e., WT, TC, and ENC) provides a hierarchical decomposition of brain tumor volume and is ubiquitously used in learning radio(geno)mics signatures (Akkus et al., 2017a; Zhou et al., 2019; Suter et al., 2020; Tixier et al., 2020).

For the discovery datasets (148 subjects in the IDH study and 118 subjects in the OS study), manual segmentation of tumor subregions (WT, TC, and ENC) was provided and confirmed by expert neuroradiologists (Menze et al., 2014). Each subject was manually annotated by one to four raters, with the same annotation protocol, and further approved/refined by experienced neuroradiologists (Menze et al., 2014). For the testing dataset (70 subjects in the IDH study and 31 subjects in the OS study), segmentation maps of tumor subregions on 3D mpMRI scans were generated using convolutional neural networks (CNNs) trained on the BraTS 2020 training cohort (369 subjects).

We employed seven state-of-the-art CNNs for brain tumor segmentation, namely, DeepMedicRes (Kamnitsas et al., 2016), Dong 2D U-Net (Dong et al., 2017), Wang 2.5D CNN (Wang et al., 2018), Isensee 3D U-Net (Isensee et al., 2019), Pereira 3D-2D U-Net (Thaha et al., 2019), HDC 3D-Net (Luo et al., 2020), and E$_1$D$_3$ 3D U-Net (Bukhari and Mohy-ud-Din, 2021). Table 3 provides a detailed summary of the architectural components and hyperparameters of the seven CNNs. We focused on segmentation architectures which can be trained with limited computational resources (GPU with 12 GB RAM) and have demonstrated superior performance in brain tumor segmentation. For instance, 3D autoencoder-regularized network (Myronenko, 2019) has a large memory footprint and is trained on a GPU with at least 32 GB RAM. Diffusion and Transformer-based models are trained on A100 GPU with 40 GB RAM (Karimi et al., 2022; Liang et al., 2022; Bieder et al., 2023; Usman Akbar et al., 2024)

Dong 2D U-Net has shown superior performance on the BraTS 2015 dataset (274 subjects). Wang 2.5D CNN and HDC 3D-Net have shown superior performance on the BraTS 2017 (487 subjects) and BraTS 2018 (542 subjects) datasets. Isensee 3D U-Net has shown superior performance on the BraTS 2018 dataset. $E_1D_3$ 3D U-Net has shown superior performance on the BraTS 2018 and the BraTS 2021 (2040 subjects) datasets. $E_1D_3$ 3D U-Net, Isensee 3D U-Net, and Wang 2.5D CNN yields overlapping subregions (WT, TC, and ENC) and DeepMedicRes, Pereira 3D-2D U-Net, Dong 2D U-Net and HDC 3D-Net provides non-overlapping subregions (PTE, NEC, and ENC) of brain tumor volume. It must be noted that one can easily obtain non-overlapping subregions from overlapping ones via set-subtraction.

DeepMedicRes, Isensee 3D U-Net, HDC 3D-Net, and $E_1D_3$ 3D U-Net employed 3D convolutions to exploit volumetric information in multiparametric images. Wang 2.5D CNN used anisotropic convolutions to tradeoff memory footprint and model complexity. More specifically, a typical $3 \times 3 \times 3$ convolution kernel is factorized into an intra-slice convolutional kernel, of size $3 \times 3 \times 1$, and an intra-slice convolutional kernel, of size $1 \times 1 \times 3$. This provides a large intra-slice receptive field and a relatively small inter-slice receptive field. Dong 2D U-Net and Pereira 3D-2D U-Net employed 2D convolutions with slice-based and patch-based input tensors respectively.

We also employed STAPLE-fusion method (Rohlfing et al., 2004) to fuse the segmentation labels obtained with DeepMedicRes, Dong 2D U-Net, Wang 2.5D CNN, Isensee 3D U-Net, Pereira 3D-2D U-Net, HDC 3D-Net, and $E_1D_3$ 3D U-Net. In a nutshell, eight segmentation schemes were employed including seven CNNs and one STAPLE-fusion method.

*2.3. Feature Extraction*

We extracted multi-regional and multi-modal radiomics features using the publicly available PyRadiomics software package (Van Griethuysen et al., 2017). More specifically, features were extracted for three overlapping tumor subregions (i.e., WT, TC, and ENC) across four 3D MRI sequences (i.e., T1, T1-Gd, T2, and FLAIR). Definition of all radiomic features were provided by Image biomarker standardization initiative (IBSI) (Zwanenburg et al., 2016). For each tumor subregion and 3D MRI sequence, 99 features were extracted including 15 shape, 17 first-order, and 67 texture features. Moreover, first-order and texture features were extracted from the original 3D MRI image, wavelet filtered images (8 wavelet bands), and Laplacian of Gaussian filtered images ($\sigma = \{1, 3\}$). In total, 11,129 radiomic features were extracted for each subject: 13 shape features $\times$ 3 tumor subregions + 2 whole brain shape features + 84 first-order and texture features $\times$ 4 channels $\times$ 11 image filtering schemes (including original images) $\times$ 3 tumor subregions. Table 4 lists the names of radiomics features extracted in this study. For the discovery dataset, radiomics features were extracted from 3D mpMRI scans using manual segmentation of (overlapping) tumor subregions provided with the BraTS 2020 dataset. For testing dataset, radiomics features were extracted using predicted segmentation maps from eight segmentation schemes (seven CNNs and one STAPLE-fusion method) elaborated in Section 2.2.

**Table 3.** A summary of architectural components and hyperparameters for seven CNNs used for automatic segmentation of brain tumor volumes.

| Network | DeepMedicRes 3D CNN | Dong 2D U-Net | Wang 2.5D CNN Three 2.5D | Isensee 3D U-Net | Pereira 3D-2D U-Net | HDC 3D-Net | EnDs 3D U-Net |
|---|---|---|---|---|---|---|---|
| Architecture | Dual-pathway (multi-scale) 3D CNN | 2D U-Net | Anisotropic CNNs (W-Net, T-Net, and E-Net) in cascade | 3D U-Net with Deep supervision | 3D-2D U-Net in cascade | 2.5D U-Net | 3D U-Net |
| Activation | P-ReLU | ReLU | P-ReLU | Leaky-ReLU (0.01) | ReLU | ReLU | Leaky-ReLU (0.01) |
| Batch size | 10 | 10 | 5 (same for three CNNs in cascade) | 2 | 3D U-Net: 1; 2D U-Net: 10 | 8 | 2 |
| Initialization | He-normal | He-normal | Truncated Normal | He-normal | He-normal | He-normal | He-normal |
| Input size / Output size | $52^3 / 9^3$ | $240^2 / 240^2$ | W-Net: $19 \times 144^2 / 11 \times 144^2$; T-Net: $19 \times 64^2 / 11 \times 64^2$; E-Net: $19 \times 64^2 / 11 \times 64^2$ | $128^3 / 128^3$ | 3D: $128^3 / 32^3$; 2D: $126^2 / 32^2$ | $128^3 / 128^3$ | $96^3 / 96^3$ |
| Learning Rate policy[a] | Polynomial Decay (batch-wise) $\eta_0 = 10^{-4}$, $\eta_{end} = 10^{-7}$, $\gamma = 1.2$ | Polynomial Decay (batch-wise) $\eta_0 = 10^{-4}$, $\eta_{end} = 10^{-7}$, $\gamma = 1.2$ | Constant ($10^{-3}$) | Polynomial decay (epoch-wise) $\eta_0 = 0.01$, $\gamma = 0.9$ | Constant ($5 \times 10^{-5}$) | Polynomial decay (epoch-wise) $\eta_0 = 10^{-3}$, $\gamma = 0.9$ | Polynomial decay (epoch-wise) $\eta_0 = 10^{-2}$, $\gamma = 0.9$ |
| Optimizer | Adam | Adam | Adam | SGD + Nesterov (0.99) | Adam | Adam (AMSGrad variant) | SGD + Nesterov (0.99) |
| Loss Function | Soft Dice | Soft Dice | Soft Dice | Soft Dice + Cross Entropy | Cross Entropy | Generalized Soft Dice | Soft Dice + Cross Entropy |
| Regularization | $L_1(10^{-6})$, $L_2(10^{-4})$ and Dropout (0.5) | – | $L_2(10^{-7})$ | $L_2(3 \times 10^{-5})$ | $L_2(10^{-5})$, Spatial Dropout 2D and 3D: (0.05) | $L_2(10^{-5})$ | $L_2(10^{-6})$ |
| Total Training iterations (Gradient-Decent updates) | 100k (1000 epochs) | 50k (100 epochs) | 20k (per-network) | 250k (1000 epochs) | 3D: 100k (100 epochs); 2D: 30k (300 epochs) | 37.35k (900 epochs) | 125k (500 epochs) |
| # Parameters | 2.8 million | 34.5 million | W-Net: 0.21 million; T-Net: 0.21 million; E-Net: 0.20 million | 31.2 million | 3D: 0.68 million; 2D: 1.64 million | 0.29 million | 34.9 million |
| Training Time[b] | ~15 hours | ~110 hours | W-Net (single-view): ~84 hours; T-Net (single-view): ~84 hours; E-Net (single-view): ~20 hours | ~101 hours | 3D: ~12.5 hours; 2D: ~23 hours | ~110 hours | ~48 hours |
| Test-time Augmentation | ✓ | ✓ | ✗ | ✓ | ✓ | ✓ | ✓ |
| Morphological Post-processing | Morphological closing, cluster thresholding | Morphological closing, cluster thresholding | ✗ | ✗ | Morphological closing, cluster thresholding | ✗ | ✓ |

NOTES: [a] For definition of variables consult Table 1 in (Bukhari and Mohy-ud-Din, 2021).
[b] Please note that training time also depends on the GPU system used for training. HDC-Net was trained on a dual-GPU system whereas remaining CNNs were trained on a single-GPU system.

**Table 4.** A list of radiomics features extracted from each subject in the discovery and testing datasets.

| Feature Type | Feature Name | Total |
|---|---|---|
| First order features | Energy, Entropy, Minimum, Maximum, 10th percentile, 90th percentile, Mean, Median, Interquartile Range, Range, MAD, rMAD, RMS, Skewness, Kurtosis, Variance, Uniformity. | 17 |
| **Shape features** | | |
| *Multi-regional features* | Mesh Volume, Surface Area, Surface Area to Volume Ratio, Sphericity, Maximum 3D Diameter, Maximum 2D Diameter (Slice), Maximum 2D Diameter (Column), Maximum 2D Diameter (Row), Major Axis Length, Minor Axis Length, Least Axis Length, Elongation, Flatness. | 13 |
| *Total Brain features* | Mesh Volume, Surface Area | 2 |
| **Texture features** | | |
| GLCM | Autocorrelation, Joint Average, Cluster Prominence, Cluster Shade, Cluster Tendency, Contrast, Correlation, Difference Average, Difference Entropy, Difference Variance, Joint Energy, Joint Entropy, IMC1, IMC2, MCC, IDMN, IDN, Inverse Variance, Maximum Probability, Sum Entropy, Sum of Squares. | 21 |
| GLRLM | Short Run Emphasis, Long Run Emphasis, Gray Level Non-Uniformity Normalized, Run Length Non-Uniformity Normalized, Run Percentage, Gray Level Variance, Run Variance, Run Entropy, Low Gray Level Run Emphasis, High Gray Level Run Emphasis, Short Run Low Gray Level Emphasis, Short Run High Gray Level Emphasis, Long Run Low Gray Level Emphasis, Long Run High Gray Level Emphasis | 14 |
| GLSZM | Small Area Emphasis, Large Area Emphasis, Gray Level Non-Uniformity Normalized, Size-Zone Non-Uniformity Normalized, Zone Percentage, Gray Level Variance, Zone Variance, Zone Entropy, Low Gray Level Zone Emphasis, High Gray Level Zone Emphasis, Small Area Low Gray Level Emphasis, Small Area High Gray Level Emphasis, Large Area Low Gray Level Emphasis, Large Area High Gray Level Emphasis. | 14 |
| GLDM | Small Dependence Emphasis, Large Dependence Emphasis, Gray Level Non-Uniformity, Dependence Non-Uniformity Normalized, Gray Level Variance, Dependence Variance, Dependence Entropy, Low Gray Level Emphasis, High Gray Level Emphasis, Small Dependence Low Gray Level Emphasis, Small Dependence High Gray Level Emphasis, Large Dependence Low Gray Level Emphasis, Large Dependence High Gray Level Emphasis. | 13 |
| NGTDM | Coarseness, Contrast, Busyness, Complexity, Strength | 5 |
| **Total Number of Features** | | 99 |

NOTES: GLCM, gray-level co-occurrence matrix; GLRLM, gray-level run length matrix; GLSZM, gray-level size zone matrix; NGTDM, neighborhood gray-tone difference matrix; MAD, Mean Absolute Deviation; rMAD, Robust Mean Absolute Deviation; RMS, Root Mean Squared; IMC1, Informational Measure of Correlation1; IMC2, Informational Measure of Correlation2; IDMN, Inverse Difference Moment Normalized; MCC, Maximal Correlation Coefficient; ID, Inverse Difference.

Radiomics features from the discovery dataset were inspected for outliers and NaNs. Outliers were identified with a scaled version of Median Absolute Deviation (MAD) as follows: $\text{MAD}_{\text{scaled}}(f_i) = c \cdot \text{median}(|f_i - \bar{f}|)$ where $c = -1/\{\sqrt{2}\Phi^{-1}3/2\}$ and $\Phi^{-1}$ is the inverse complementary error function (Leys et al., 2013). Every feature value $f_i^k > 3\ \text{MAD}_{\text{scaled}}(f_i)$ was labeled as an outlier and replaced by the mean of the remaining feature values. The NaNs in each feature vector ($f_i$) were also replaced by the same mean value. This was followed by *z*-score normalization of the feature vectors ($f_i$). Radiomics features from the testing dataset were independently corrected for NaNs followed by *z*-score normalization. For each feature vector in the testing dataset, NaNs were replaced by the mean values used (to replace outliers and NaNs) in the discovery dataset and *z*-score normalization was performed using the mean and standard deviation computed on the discovery dataset.

*2.4. Stability Analysis*

We used overall concordance correlation coefficient (OCCC) (Barnhart et al., 2002) to quantitatively measure robustness of radiomics features across seven (independent) segmentation schemes i.e., seven CNNs mentioned in Section 2.2. OCCC is a frequently used measure of agreement of radiomics features across multiple raters (Kalpathy-Cramer et al., 2016b; Li et al., 2019) and is computed as a weighted average of all pairwise concordance correlation coefficients (Lin, 1989). OCCC incorporates both the degree of agreement and disagreement by assigning higher weights to pairs of raters whose measurements have higher variances and larger mean differences (Barnhart et al., 2002).

In the spirit of domain adaptation (French et al., 2017), we selected a subset of radiomics features for training which are robust to variations in segmentations on the testing dataset. More specifically, the automatically generated brain tumor segmentations, from seven CNNs, on testing dataset were recognized as seven independent raters for reliability analysis. We used OCCC to select a subset of robust features from the original pool of 11,129 radiomics features. A radiomics feature was said to be robust if $\text{OCCC} \geq 0.95$.

Reliability scores (OCCC), computed on testing dataset, were only used to identify a pool of robust features (i.e., feature names only) for training a radio(geno)mics model, on the discovery dataset, which exhibits strong generalizability on the novel dataset. This process is referred to as stability filtering in the manuscript.

*2.5. Feature Selection*

We employed two feature selection (FS) methods to select an optimal subset of informative and discriminatory radiomic features. It is done in two stages. In stage 1, noninformative features with median absolute deviation (MAD) of zero were eliminated. In stage 2, a subset of discriminatory features was selected using one of the following FS methods:

- Minimum Redundancy, Maximum Relevance (MRMR) (Ding and Peng, 2003): MRMR selects a subset of minimally redundant features, $\{F_i\}$, quantified by a small average pairwise Pearson correlation $\text{corr}(F_i, F_j)$ for all $1 \leq i, j \leq |F|$, which are strongly associated with response variables ($Y$), quantified by a large $\mathcal{F}$ statistic $\mathcal{F}(F_i, Y)$. We preferred a statistical formulation over an information-theoretic one as recommended in several studies (Ding and Peng, 2005; Radovic et al., 2017; Zhao et al., 2019; Bera et al., 2022).
- Recursive Feature Elimination with SVM (RFE-SVM) (Rakotomamonjy, 2003): RFE begins by training an SVM classifier with the complete set of features and eliminates the one with the lowest feature importance score. This process is repeated on the reduced set of features until the required number of features are reached. RFE is superior to forward feature selection approach as every feature is considered in the selection process.

We also quantified the predictive power of each radiomics feature by computing univariate AUC (uAUC) as follows. A single radiomics feature, at a time, was used to train a Random Forest classifier for classification task (i.e., IDH mutation study and OS classification study). The uAUC of each feature was an average over 100 iterations of randomized and stratified splitting of the discovery dataset ($70\% - 30\%$ split).

*2.6. Pipeline for Model Training*

We performed two sets of experiments to establish the efficacy of training a radio(geno)mics model with (a subset of) features that are robust to variability in segmentation on the novel (unseen) testing dataset. Our hypothesis is that incorporating robust features in model training will enhance generalizability of radio(geno)mics model on novel datasets. The original set of radiomics features (11,129 features) was augmented with a clinical feature, i.e., Age, before subsequent feature selection and model training.

- Experiment 1 (Without Stability Filtering): In the first experiment, the original set of features was reduced to a subset of informative (via MAD filtering) and discriminatory features (via one of the two feature selection methods outlined in Section 2.5). The obtained subset of discriminatory features was used to train fifty ($n = 50$) random forest classifiers with random initialization.
- Experiment 2 (With Stability Filtering): In the second experiment, the original set of features was first reduced to a subset of robust features (via stability filtering as outlined in Section 2.4) followed by further reduction to a subset of informative and discriminatory features. The obtained subset of stable and discriminatory features was used to train fifty ($n = 50$) random forest classifiers with random initialization.

Hyperparameters of the random forest classifier were set as follows: no_of_estimators = 200, max_features = auto, class_weight = balanced, criterion = gini. Hyperparameters for the two feature selection methods were set as follows:

MRMR  – n_selected_features = $N$
RFE-SVM  – n_selected_features = $N$, kernel = linear, and step = 1

Class imbalance in the IDH discovery dataset (Shannon Entropy = 0.96) was addressed using Synthetic Minority Oversampling technique (SMOTE) with five neighbors (Chawla et al., 2002). The discovery dataset for OS classification task was quite balanced (Shannon Entropy = 0.99) across the three survival classes and, hence, SMOTE was not applied. For model training, we explored optimal subset of features of varying cardinalities.

For inference on the testing dataset, a soft voting method was adopted to unify the outputs of 50 random forest classifiers (with uniform weighting scheme) and generate a single prediction (IDH: mutation vs wild type, OS: short-term vs medium-term vs long-term survival). Figure 2 shows a schematic diagram of the proposed experiments for training and inference.

*2.7. Evaluation Criteria*

Performance of the eight segmentation schemes (seven CNNs and one STAPLE-fusion method) was quantified using Dice Similarity Coefficient (DSC) (Dice, 1945) and Hausdorff distance metric (HD-95) (Huttenlocher et al., 1992). The eight segmentation schemes were ranked based on the Final Ranking Score (FRS) and statistical significance (of ranking) was calculated using random permutation test (Bukhari and Mohy-ud-Din, 2021). Predictive performance of radio(geno)mics models was quantified using the area under the receiver operating curve (AUC). Statistical comparisons of AUC were performed using Fast DeLong method (Sun and Xu, 2014). Stability of the radio(geno)mics models was quantified with relative standard deviation (RSD) calculated as a ratio of standard deviation to the mean of AUC. A lower value of RSD corresponds to higher stability of the radio(geno)mics model. Statistical analysis of demographic data (in Table 1a and Table 1b) was performed using the student $t$-test. A $p < 0.05$ was considered statistically significant.

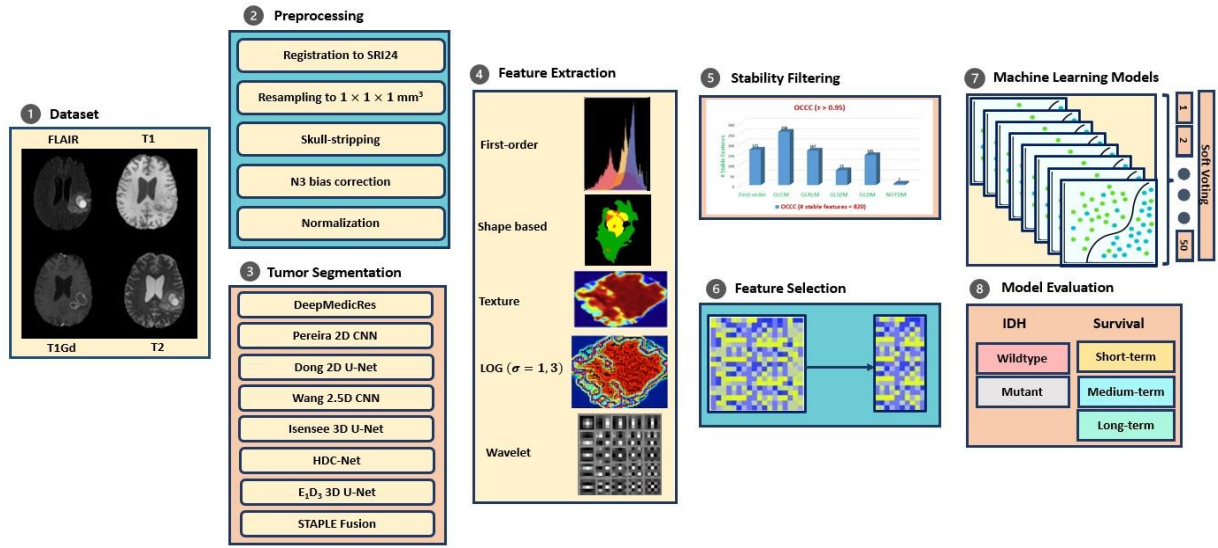

**Figure 2.** *Workflow for training and inference*: (1) Acquisition of 3D mpMRI scans including T1, T1-Gd, T2, and FLAIR sequences. (2) Preprocessing of 3D mpMRI scans; described in Section 2.1.4. (3) Multi-regional segmentation maps obtained with fully automatic deep segmentation methods; described in Section 2.2. (4) Extraction of radiomics features using the PyRadiomics package; described in Section 2.3. (5) Identifying radiomics features stable to variability in multi-regional segmentation maps obtained with fully automatic deep segmentation methods; described in Section 2.4. (6) Selection of discriminatory features with RFE-SVM or MRMR; described in Section 2.5. (7) Learning radiomics (and radiogenomics) signature with Random Forest classifier trained with (stable) discriminatory features; described in Section 2.6. (8) Evaluation of radiomics (and radiogenomics) models; described in Section 2.6 and Section 2.7.

*2.8. System Specification*

All results were generated using python open-source packages: scikit-feature (García et al., 2015), sklearn (Barupal and Fiehn, 2019), NumPy (Harris et al., 2020), Pandas (McKinney, 2010), scipy (Kiyohara et al., 2015), imblearn (Kiyohara et al., 2015), matplotlib (Hunter, 2007), PyRadiomics[b] (Van Griethuysen et al., 2017), ANTs (Avants et al., 2009), STAPLE fusion[c] (Rohlfing et al., 2004), OCCC[d] (Barnhart et al., 2002b), and DeLong method[e] (Delong and Carolina, 1988; Sun and Xu, 2014).

## 3. Results

*3.1. Clinical Characteristics*

*3.1.1. IDH prediction study.* We analyzed data from a multicenter cohort of 218 subjects (see Table 2a). Eight out of nine institutions were represented in the discovery and testing datasets. Fondazione IRCCS Instituto Neuroligico C. Besta (FINCB) only contributed three subjects to the testing dataset. Moreover, 218 subjects were quite unevenly distributed across the nine institutions with Henry Ford Hospital contributing ∼41% of the total subjects and Duke, Emory, and FINCB contributing < 3.5% each of the total subjects.

---

[b] https://pyradiomics.readthedocs.io
[c] https://github.com/FETS-AI/LabelFusion
[d] https://rdrr.io/cran/epiR/src/R/epi.occc.R
[e] https://github.com/yandexdataschool/roc_comparison

Table 1a provides a summary of subjects' characteristics in the discovery and testing datasets. The discovery dataset (148 subjects) comprised of 76 males and 71 females with a mean age of 51 years. Histology analysis revealed that 21 subjects were diagnosed with Astrocytoma, 18 subjects with Oligoastrocytoma, 25 subjects with Oligodendroglioma, and 83 subjects with Glioblastoma. Molecular testing revealed that 57 subjects had IDH mutation (IDHmutant) and 91 subjects had IDH-wild type (IDHwt). The testing dataset (70 subjects) comprised of 40 males and 30 females with a mean age of 54 years. Histology analysis revealed that 11 subjects were diagnosed with Astrocytoma, 10 subjects with Oligoastrocytoma, 21 subjects with Oligodendroglioma, and 28 subjects with Glioblastoma. Molecular testing revealed that 32 subjects had IDH mutation (IDHmutant) and 38 subjects had IDH-wild type (IDHwt). No statistical difference was found in age ($p = 0.20$), gender ($p = 0.54$), tumor grade ($p = 0.06$), histology outcome ($p = 0.09$), and IDH status ($p = 0.38$) between the discovery dataset and the testing dataset. However, KPS was significantly higher ($p = 0.04$) in the discovery dataset (mean KPS = $83.9 \pm 12$) in comparison to the testing dataset (mean KPS = $77.8 \pm 17.2$).

*3.1.2. Survival Study.* Subjects in the training and testing datasets (149) were pooled from nine institutions (Table 2b). Overall, 80% of subjects in the training dataset was from the University of Pennsylvania (UPenn). Henry Ford Hospital contributed eleven subjects to the testing dataset and only one subject to the training dataset.

Table 1b displays clinical characteristics of the discovery and testing datasets. The mean age of subjects in the discovery dataset (118 subjects) was 61.9 years. Discovery dataset was balanced across the three survival groups, i.e., short-term (42 subjects), medium-term (30 subjects), and long-term (46 subjects) survivors. The mean age of subjects in the testing dataset was 58 years. Testing dataset (31 subjects) had a sparse presence of medium-term survivors – only 3 subjects out of 31. The mean OS (in days) for the discovery dataset and testing dataset were 446 days and 293 days respectively. No statistical difference was found in age between the discovery dataset and the testing dataset ($p = 0.252$). No statistical difference was found in OS days between the discovery dataset and testing dataset ($p = 0.40$).

**Table 5.** Performance of eight segmentation schemes, including seven CNNs and one STAPLE-fusion method, on testing dataset (125 subjects). **Bold font** indicates best scores for overlapping subregions.

| Segmentation Network | Dice Similarity Coefficient (%) | | | Hausdorff-95 Distance (mm) | | | FRS |
|---|---|---|---|---|---|---|---|
| | WT | TC | ENC | WT | TC | ENC | |
| DeepMedicRes 3D CNN | $88.7 \pm 12.3$ | $78.1 \pm 25.8$ | $71.6 \pm 31.6$ | $8.97 \pm 17.0$ | $17.7 \pm 57.3$ | $32.3 \pm 96.0$ | $5^{**}$ |
| Dong 2D U-Net | $89.6 \pm 7.2$ | $77.7 \pm 23.7$ | $71.0 \pm 29.4$ | $5.45 \pm 7.7$ | $11.4 \pm 34.5$ | $37.3 \pm 105.0$ | $7^{**}$ |
| Wang 2.5D CNN | $88.1 \pm 13.0$ | $77.4 \pm 25.3$ | $75.2 \pm 28.3$ | $11.1 \pm 20.8$ | $13.67 \pm 36.4$ | $29.0 \pm 91.2$ | $6^{**}$ |
| Isensee 3D U-Net | $90.5 \pm 8.1$ | $\mathbf{84.5 \pm 16.4}$ | $76.9 \pm 27.9$ | $\mathbf{4.41 \pm 5.99}$ | $\mathbf{8.65 \pm 34.4}$ | $32.6 \pm 100.9$ | 1 |
| Pereira 3D-2D U-Net | $87.7 \pm 11.9$ | $69.5 \pm 30.2$ | $67.0 \pm 32.1$ | $13.9 \pm 23.3$ | $22.85 \pm 51.0$ | $45.5 \pm 108.8$ | $8^{**}$ |
| HDC 3D-Net | $89.6 \pm 10.3$ | $83.1 \pm 18.5$ | $\mathbf{77.5 \pm 27.2}$ | $7.5 \pm 33.5$ | $12.4 \pm 47.5$ | $32.3 \pm 100.9$ | $3^{**}$ |
| $E_1D_3$ 3D U-Net | $\mathbf{90.6 \pm 6.4}$ | $82.7 \pm 19.9$ | $76.4 \pm 27.6$ | $5.8 \pm 10.2$ | $10.8 \pm 35.9$ | $\mathbf{22.9 \pm 79.9}$ | $4^{**}$ |
| STAPLE Fusion | $90.4 \pm 7.4$ | $82.9 \pm 19.3$ | $74.8 \pm 28.8$ | $5.3 \pm 9.4$ | $12.3 \pm 47.4$ | $30.6 \pm 96.1$ | 2 |

NOTE: $^{**}$indicates that the segmentation network is ranked significantly lower ($p < 0.001$) in comparison to the top ranked method Isensee 3D U-Net (FRS = 1).

*3.2. Automatic Brain Tumor Segmentation*

Table 5 summarizes the performance of eight segmentation schemes (seven CNNs and one STAPLE-fusion method) for the testing dataset (125 subjects). In terms of FRS, Isensee 3D U-Net and STAPLE fusion method were ranked first and second, respectively, with no significant difference ($p = 0.49$). However, STAPLE fusion method and Isensee 3D U-Net were ranked significantly higher ($p < 0.001$) in comparison to the remaining six CNNs for brain tumor segmentation. In terms of DSC metric, which quantifies degree of overlap with the (ground-truth) manual segmentations, $E_1D_3$ 3D U-Net provided best segmentation performance for WT ($DSC = 90.6 \pm 6.4\%$), Isensee 3D U-Net provided best segmentation performance for TC ($DSC = 84.5 \pm 16.4\%$), and HDC 3D-Net provided best segmentation performance for ENC ($DSC = 77.5 \pm 27.2\%$). In terms of HD-95 metric, Isensee 3D U-Net reported lowest voxel-wise segmentation errors for WT ($HD\text{-}95 = 4.41 \pm 5.99$ mm) and TC ($HD\text{-}95 = 8.65 \pm 34.4$ mm) and $E_1D_3$ 3D U-Net reported lowest voxel-wise segmentation error for ENC ($HD\text{-}95 = 22.9 \pm 79.9$ mm). Pereira 3D-2D U-Net reported lowest segmentation performance, in terms of DSC and HD-95 metrics, for all tumor subregions and, hence, was ranked last (in terms of FRS).

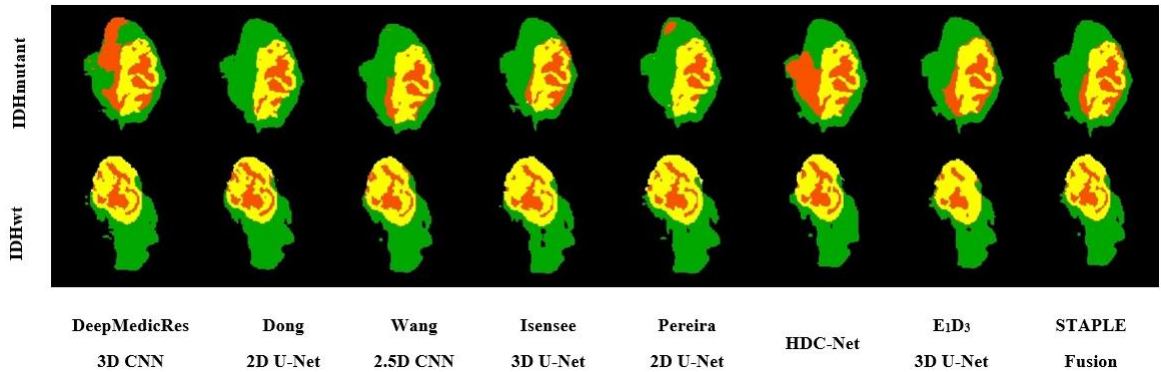

**Figure 3.** *Multi-class segmentation maps obtained with eight segmentation schemes for two subjects – IDHmutant and IDHwt*: Green, Yellow and Orange represent peritumoral edema, enhancing tumor, and non-enhancing tumor respectively. It must be noted that ground truth segmentation maps are not available for the testing dataset (125 subjects). Label legend: **Peritumoral Edema**, **Enhancing Core**, **Non-enhancing Core.**

Figure 3 shows the predicted multi-class segmentation maps obtained with eight segmentation schemes for two subjects – IDHmutant and IDHwt. Likewise, Figure 4 shows the predicted multi-class segmentation maps obtained with eight segmentation schemes for three subjects – short-term, medium-term, and long-term survivor.

*3.3. Stability of Radiomics Features*

Firstly, from the original set of 11,129 features, mined from the validation cohort of BraTS 2020 dataset (125 subjects), we removed noninformative features (identified with MAD = 0) and features not influenced by automatic segmentation, which includes Total Brain features (see Table 4) and Age. This reduced the original set to 11045 radiomics features. Secondly, we extracted a subset of radiomics features (from a pool of 11045 features) stable to variations in segmentations across seven CNNs (see Section 2.2). More specifically, a subset of highly stable radiomics features was identified with an OCCC ≥ 0.95 across seven segmentation schemes. Figure 5 summarizes the number of stable features obtained for each feature category by *stability filtering* using OCCC.

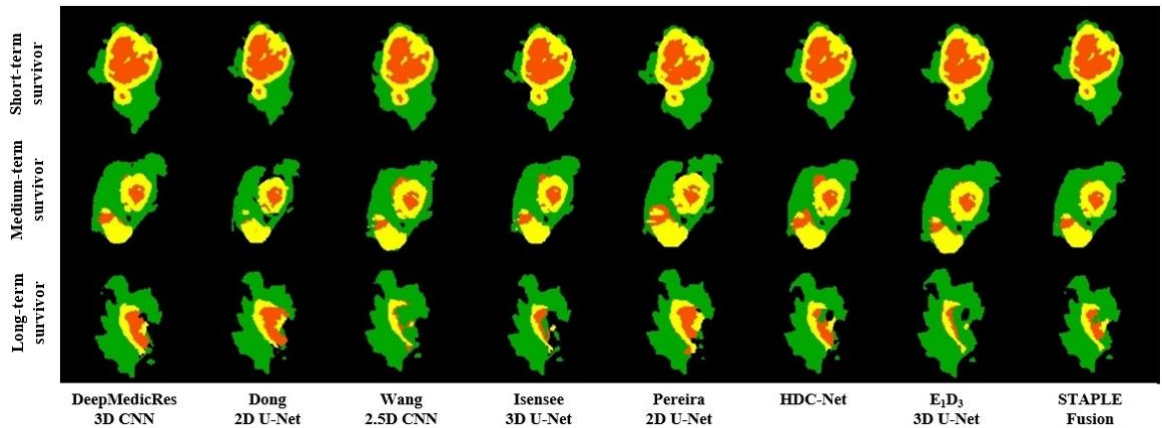

**Figure 4.** *Multi-class segmentation maps obtained with eight segmentation schemes for three subjects – short-term, medium-term and long-term survivors*: Green, Yellow and Orange represent peritumoral edema, enhancing tumor, and non-enhancing tumor respectively. It must be noted that ground truth segmentation maps are not available for the testing dataset (125 subjects). Label legend**: Peritumoral Edema, Enhancing Core, Non-enhancing Core.**

Stability filtering with OCCC ($\tau \geq 0.95$) yielded 820 stable features with the following statistics: *Feature Category* – 171 first-order features, 649 texture features, *MRI Sequence* – 151 features from FLAIR sequence, 294 features from T1-Gd sequence, 230 features from T1 sequence, and 145 features from T2 sequence, and *Tumor Subregion* – 791 features from WT (96.5%) and 29 features from TC (3.5%). No features were selected from the ENC tumor subregion. The obtained set of highly stable radiomics features (820) were augmented with Total Brain features (2) and Age to yield a subset of stable features (823) for prospective feature selection.

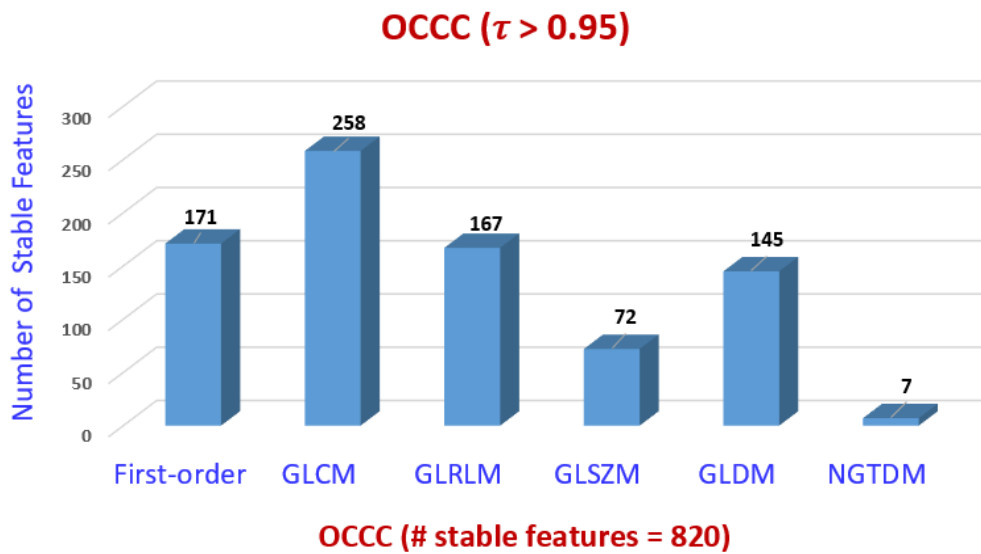

**Figure 5.** *Stability filtering of radiomics features*: Distribution of highly stable radiomics features across different feature categories.

**Table 6.** A summary of discriminatory features selected to build the radiogenomics model for IDH mutation study.

| IDH Study | MRMR | | | | Without Stability Filtering | |
|---|---|---|---|---|---|---|
| No. | Selected feature name | Type | Region | Modality | uAUC ± std | OCCC |
| 1 | Root Mean Squared | First-order | TC | T1-Gd | 0.85 ± 0.05 | 0.76 |
| 2 | Uniformity | First-order | EN | T2 | 0.75 ± 0.08 | 0.79 |
| 3 | Cluster Prominence | GLCM | EN | T1-Gd | 0.51 ± 0.06 | 0.60 |
| 4 | Mean | First-order | TC | T1-Gd | 0.84 ± 0.05 | 0.72 |
| 5 | High Gray Level Emphasis | GLDM | TC | T1-Gd | 0.84 ± 0.05 | 0.89 |
| 6 | Mean | First-order | TC | T1-Gd | 0.83 ± 0.05 | 0.72 |
| 7 | Skewness | First-order | TC | T1-Gd | 0.84 ± 0.05 | 0.82 |
| 8 | Short Run Low Gray Level Emphasis | GLRLM | WT | FLAIR | 0.53 ± 0.07 | 0.69 |
| 9 | Auto correlation | GLCM | TC | T1-Gd | 0.85 ± 0.05 | 0.89 |
| 10 | Joint Average | GLCM | TC | T1-Gd | 0.81 ± 0.05 | 0.80 |

| IDH Study | RFE-SVM | | | | Without Stability Filtering | |
|---|---|---|---|---|---|---|
| No. | Selected feature name | Type | Region | Modality | uAUC ± std | OCCC |
| 1 | 90th Percentile | First-order | EN | T1 | 0.68 ± 0.06 | 0.78 |
| 2 | Auto Correlation | GLCM | EN | T2 | 0.79 ± 0.05 | 0.75 |
| 3 | 10th Percentile | First-order | EN | T2 | 0.80 ± 0.06 | 0.78 |
| 4 | Contrast | NGTDM | EN | T2 | 0.70 ± 0.07 | 0.73 |
| 5 | Small Area Low Gray Level Emphasis | GLSZM | TC | T1-Gd | 0.57 ± 0.07 | 0.68 |
| 6 | Busyness | NGTDM | TC | T1-Gd | 0.62 ± 0.06 | 0.65 |
| 7 | Cluster Prominence | GLCM | TC | T1-Gd | 0.60 ± 0.07 | 0.84 |
| 8 | Low Gray Level Zone Emphasis | GLSZM | WT | FLAIR | 0.66 ± 0.06 | 0.68 |
| 9 | Dependence Variance | GLDM | WT | FLAIR | 0.56 ± 0.06 | 0.80 |
| 10 | Age | Clinical | - | - | 0.71 ± 0.07 | 1 |

| IDH Study | MRMR | | | | With Stability Filtering | |
|---|---|---|---|---|---|---|
| No. | Selected feature name | Type | Region | Modality | uAUC ± std | OCCC |
| 1 | Large Dependence High Gray Level Emphasis | GLDM | WT | T1-Gd | 0.86 ± 0.06 | 0.95 |
| 2 | Entropy | First-order | WT | T1 | 0.52 ± 0.07 | 0.98 |
| 3 | Large Dependence Emphasis | GLDM | WT | T1 | 0.60 ± 0.07 | 0.96 |
| 4 | Gray Level Non-Uniformity Normalized | GLRLM | WT | T2 | 0.73 ± 0.07 | 0.97 |
| 5 | Low Gray Level Zone Emphasis | GLSZM | WT | T1-Gd | 0.75 ± 0.06 | 0.95 |
| 6 | Long Run High Gray Level Emphasis | GLRLM | WT | T1-Gd | 0.85 ± 0.05 | 0.96 |
| 7 | Zone Percentage | GLSZM | WT | T2 | 0.75 ± 0.06 | 0.97 |
| 8 | Small Area Low Gray Level Emphasis | GLSZM | WT | T1-Gd | 0.80 ± 0.05 | 0.96 |
| 9 | Busyness | NGTDM | WT | T1-Gd | 0.73 ± 0.07 | 0.96 |
| 10 | Age | Clinical | - | - | 0.71 ± 0.07 | 1 |

| IDH Study | RFE-SVM | | | | With Stability Filtering | |
|---|---|---|---|---|---|---|
| No. | Selected feature name | Type | Region | Modality | uAUC ± std | OCCC |
| 1 | IMC2 | GLCM | WT | FLAIR | 0.61 ± 0.06 | 0.97 |
| 2 | Run Entropy | GLRLM | WT | FLAIR | 0.55 ± 0.07 | 0.97 |
| 3 | Dependence Entropy | GLDM | WT | T1-Gd | 0.57 ± 0.07 | 0.96 |
| 4 | Large Dependence High Gray Level Emphasis | GLDM | WT | T1-Gd | 0.86 ± 0.06 | 0.95 |
| 5 | Long Run Emphasis | GLRLM | WT | T1-Gd | 0.60 ± 0.06 | 0.99 |
| 6 | Difference Average | GLCM | WT | T1-Gd | 0.56 ± 0.09 | 0.97 |
| 7 | Variance | First-order | WT | T1 | 0.69 ± 0.06 | 0.95 |
| 8 | Dependence Variance | GLDM | WT | T1 | 0.51 ± 0.06 | 0.98 |
| 9 | Run Entropy | GLRLM | WT | T2 | 0.67 ± 0.07 | 0.97 |
| 10 | Run Entropy | GLRLM | WT | T2 | 0.56 ± 0.06 | 0.97 |

*3.4. IDH Study*

*3.4.1. Feature selection*. We employed one of the two feature selection methods – MRMR and RFE-SVM – to obtain an optimal subset of discriminatory features for the underlying radiogenomics task. The size (or cardinality) of the optimal subset of features was controlled by *a priori* setting the number of features (*N*) in MRMR and RFE-SVM. Details of the selected features are presented in Table 6.

*Without Stability Filtering:* The original set of radiomics and clinical features (11,130), from the discovery dataset, was first reduced to an informative subset of features via MAD filtering (11,060 features) followed by further reduction to a subset of discriminatory features:

(a) MRMR: We found that superior predictive performance was obtained for an optimal subset of 10 features including 5 first-order and 5 texture features. Shape features were not selected. The statistics of the selected features were: mean OCCC = 0.77 ± 0.09 and mean uAUC = 0.77 ± 0.13. No stable features (OCCC ≥ 0.95) were selected.

(b) RFE-SVM: We found that superior predictive performance was obtained for an optimal subset of 10 features including Age, 2 first-order, and 7 texture features. Shape features were not selected. The statistics of the selected features were: mean OCCC = 0.77 ± 0.01 and mean uAUC = 0.67 ± 0.08. No stable features (OCCC ≥ 0.95) were selected.

*With Stability Filtering:* Post identification of an augmented subset of stable radiomics features (823) on the validation cohort of BraTS 2020 dataset (125 subjects), as elaborated in Section 3.3, the corresponding feature names were used to extract radiomics features from the discovery dataset. All non-informative features were removed before stability filtering via MAD filtering.

(a) MRMR: We found that superior predictive performance was obtained for an optimal subset of 10 features including Age, 1 first-order, and 8 texture features. The statistics of the selected features were: mean OCCC = 0.97 ± 0.01 and mean uAUC = 0.73 ± 0.01.

(b) RFE-SVM: We found that superior predictive performance was obtained for an optimal subset of 10 features including 1 first-order and 9 texture features. The statistics of the selected features were: mean OCCC = 0.97 ± 0.01 and mean uAUC = 0.62 ± 0.01.

*Important Note:* It must be noted that the validation cohort of BraTS 2020 dataset was only used to identify radiomics features stable to variations in segmentations independently obtained with seven CNNs. Feature names associated with the subset of stable features were used to mine radiomics features from the discovery dataset which were ultimately used for subsequent feature selection and model training. *Hence, no information leakage occurred between model training and inference phases.*

**Table 7.** Predictive performance of radiogenomics models, quantified with AUROC, for IDH classification task. Each radiogenomics model corresponds to a specific combination of feature selection method, deep segmentation network, and stability filtering. For inference on the testing dataset (70 subjects), a soft voting method was adopted to unify the outputs of 50 random forest classifiers (with uniform weighting scheme) and generate a single prediction (IDHmutant vs IDHwt). Bold font indicates best predictive performance (higher AUROC) for each radiogenomics model.

| Feature Selection Method | Stability filtering | Number of features | DeepMedicRes 3D CNN | Dong 2D U-Net | Wang 2.5D U-Net | Isensee 3D U-Net | Pereira 3D-2D U-Net | HDC 3D-Net | $E_1D_3$ 3D U-Net | STAPLE Fusion |
|---|---|---|---|---|---|---|---|---|---|---|
| MRMR | True | 10 | **0.878** | 0.851 | **0.895** | 0.874 | **0.876** | **0.865** | **0.876** | **0.887** |
|  | False | 10 | 0.840 | **0.865** | 0.850 | **0.876** | 0.847 | 0.838 | 0.863 | 0.823 |
| RFE-SVM | True | 10 | **0.943*** | **0.930**** | **0.947**** | **0.935*** | **0.944**** | **0.935*** | **0.943*** | **0.932*** |
|  | False | 10 | 0.820 | 0.795 | 0.806 | 0.830 | 0.770 | 0.833 | 0.820 | 0.821 |

NOTES: *represents statistically significant improvement ($p$-value < 0.05) and **represents statistically highly significant improvement ($p$-value < 0.005)

*3.4.2. Performance evaluation.* Table 7 summarizes the predictive performance of two feature selection methods, with and without *stability filtering* and across eight segmentation schemes (seven CNNs and one STAPLE-fusion method), using AUROC. Statistical comparisons of AUROCs were made using the fast DeLong method and robustness of radiogenomics models, across eight segmentation schemes, was quantified with Relative Standard Deviation (RSD) of AUROCs.

*Without Stability Filtering:* The average predictive performance of the two feature selection methods, across eight segmentation schemes, were as follows: MRMR – AUROC $0.85 \pm 0.02$ and RFE-SVM – AUROC $0.81 \pm 0.02$. The stability of the models was 1.92% with MRMR and 2.28% with RFE-SVM as measured with RSD, across eight segmentation methods.

*With Stability Filtering:* The average predictive performance of the two feature selection methods, across eight segmentation schemes, were as follows: MRMR – AUROC $0.88 \pm 0.01$ and RFE-SVM – AUROC $0.94 \pm 0.006$. The stability of the models was 1.43% with MRMR and 0.64% with RFE-SVM as measured with RSD, across eight segmentation methods.

Stability filtering, followed by RFE-SVM guided feature selection, significantly improved ($p < 0.05$) predictive performance across all segmentation networks. For segmentation networks ranked at the bottom, in terms of FRS (see Table 5), namely Wang 2.5D U-Net, Dong 2D U-Net, and Pereira 3D-2D U-Net, stability filtering brought statistically highly significant improvements in predictive performance ($p < 0.005$).

Stability filtering, followed by MRMR guided feature selection, did report higher AUROC for six out of eight segmentation networks but the improvement in predictive performance was not statistically significant ($p \in [0.09, 0.73]$). For Isensee 3D U-Net and Dong 2D U-Net segmentation networks, AUROC decreased by 0.2% ($p = 0.93$) and 1.4% ($p = 0.73$) respectively.

Stability filtering with RFE-SVM guided feature selection, as opposed to its MRMR counterpart, reported improved predictive performance across all segmentation schemes. More specifically, statistically significant improvements in predictive performance were obtained with DeepMedicRes 3D CNN, Dong 2D U-Net, Pereira 3D-2D U-Net, and HDC 3D-Net ($p < 0.05$).

**Table 8.** A summary of discriminatory features selected to build the radiomics model for OS classification study.

| No. | Selected feature name | Type | Region | Modality | uAUC ± std | OCCC |
|---|---|---|---|---|---|---|
| | **OS Classification** | **MRMR** | | | **Without Stability Filtering** | |
| 1 | Low Gray Level Run Emphasis | GLRLM | EN | T1-Gd | 0.57 ± 0.05 | 0.26 |
| 2 | Short Run High Gray Level Emphasis | GLRLM | WT | T2 | 0.56 ± 0.05 | 0.79 |
| 3 | MCC | GLCM | EN | FLAIR | 0.60 ± 0.06 | 0.68 |
| 4 | Skewness | First-order | TC | T1 | 0.51 ± 0.05 | 0.36 |
| 5 | Maximum Probability | GLCM | WT | T2 | 0.62 ± 0.06 | 0.85 |
| 6 | Kurtosis | First-order | EN | T2 | 0.60 ± 0.06 | 0.72 |
| 7 | IMC2 | GLCM | TC | T1-Gd | 0.62 ± 0.05 | 0.79 |
| 8 | Run Entropy | GLRLM | TC | FLAIR | 0.49 ± 0.06 | 0.86 |
| 9 | Short Run Low Gray Level Emphasis | GLRLM | EN | T1-Gd | 0.54 ± 0.05 | 0.27 |
| 10 | Age | Clinical | - | - | 0.54 ± 0.06 | 1 |
| | **OS Classification** | **MRMR** | | | **With Stability Filtering** | |
| 1 | Contrast | GLCM | WT | T1-Gd | 0.51 ± 0.06 | 0.98 |
| 2 | Zone Entropy | GLSZM | WT | T1 | 0.50 ± 0.06 | 0.97 |
| 3 | Run Variance | GLRLM | WT | T2 | 0.51 ± 0.05 | 0.96 |
| 4 | Uniformity | First-order | TC | T1 | 0.57 ± 0.05 | 0.95 |
| 5 | Long Run High Gray Level Emphasis | GLRLM | WT | T1-Gd | 0.57 ± 0.06 | 0.96 |
| 6 | Long Run Emphasis | GLRLM | WT | T2 | 0.51 ± 0.05 | 0.97 |
| 7 | Dependence Entropy | GLDM | WT | T1 | 0.53 ± 0.05 | 0.99 |
| 8 | Robust Mean Absolute Deviation | First-order | WT | T1-Gd | 0.52 ± 0.04 | 0.95 |
| 9 | Inverse Variance | GLCM | WT | T1-Gd | 0.45 ± 0.06 | 0.95 |
| 10 | Age | Clinical | - | - | 0.54 ± 0.06 | 1 |
| | **OS Classification** | **RFE-SVM** | | | **Without Stability Filtering** | |
| 1 | Difference Variance | GLCM | EN | T1 | 0.54 ± 0.06 | 0.76 |
| 2 | Low Gray Level Run Emphasis | GLRLM | EN | T1-Gd | 0.57 ± 0.05 | 0.26 |
| 3 | Short Run Low Gray Level Emphasis | GLRLM | EN | T1-Gd | 0.54 ± 0.05 | 0.27 |
| 4 | Long Run Low Gray Level Emphasis | GLRLM | EN | T2 | 0.49 ± 0.05 | 0.56 |
| 5 | Small Dependence High Gray Level Emphasis | GLDM | TC | T1 | 0.54 ± 0.06 | 0.82 |
| 6 | Run Variance | GLRLM | TC | T1 | 0.52 ± 0.06 | 0.90 |
| 7 | Kurtosis | First-order | TC | T2 | 0.52 ± 0.06 | 0.67 |
| 8 | Robust Mean Absolute Deviation | First-order | TC | T2 | 0.58 ± 0.06 | 0.84 |
| 9 | 10th Percentile | First-order | WT | FLAIR | 0.54 ± 0.06 | 0.94 |
| 10 | Long Run Low Gray Level Emphasis | GLRLM | WT | T1 | 0.54 ± 0.07 | 0.53 |
| | **OS Classification** | **RFE-SVM** | | | **With Stability Filtering** | |
| 1 | Joint Energy | GLCM | WT | FLAIR | 0.47 ± 0.06 | 0.95 |
| 2 | Large Dependence Emphasis | GLDM | WT | FLAIR | 0.48 ± 0.06 | 0.95 |
| 3 | IMC2 | GLCM | WT | FLAIR | 0.45 ± 0.05 | 0.96 |
| 4 | Joint Energy | GLCM | WT | FLAIR | 0.49 ± 0.05 | 0.97 |
| 5 | Small Area High Gray Level Emphasis | GLSZM | WT | T1-Gd | 0.54 ± 0.05 | 0.97 |
| 6 | Uniformity | First-order | TC | T1 | 0.57 ± 0.05 | 0.96 |
| 7 | Run Entropy | GLRLM | WT | T1 | 0.42 ± 0.06 | 0.99 |
| 8 | Dependence Variance | GLDM | WT | T2 | 0.55 ± 0.05 | 0.98 |
| 9 | Large Dependence High Gray Level Emphasis | GLDM | WT | T2 | 0.49 ± 0.06 | 0.96 |
| 10 | Age | Clinical | - | - | 0.54 ± 0.06 | 1 |

*3.5. Overall Survival Classification*

*3.5.1. Feature selection.* We employed one of the two feature selection methods – MRMR and RFE-SVM – to obtain an optimal subset of discriminatory features for the underlying radiomics task. The size (or cardinality) of the optimal subset of features was controlled by *a priori* setting the number of features (*N*) in MRMR and RFE-SVM. Details of the selected features are presented in Table 8.

*Without Stability Filtering:* The original set of radiomics and clinical features (11130), from the discovery dataset, was first reduced to an informative subset of features via MAD filtering (11058 features) followed by further reduction to a subset of discriminatory features.

(a) MRMR: We found that superior predictive performance was obtained for an optimal subset of 10 features including Age, 2 first-order, and 7 texture. Shape features were not selected. The statistics of the selected features were: mean OCCC = $0.66 \pm 0.3$ and mean uAUC = $0.56 \pm 0.04$. No stable features (OCCC $\geq$ 0.95) were selected with MRMR.

(b) RFE-SVM: We found that superior predictive performance was obtained for an optimal subset of 10 features including 3 first-order, and 7 texture features. Shape features were not selected. The statistics of the selected features were: mean OCCC = $0.66 \pm 0.2$ and mean uAUC = $0.54 \pm 0.02$. No stable features (OCCC $\geq$ 0.95) were selected with RFE-SVM.

*With Stability Filtering:* Post identification of an augmented subset of stable radiomics features (823) on the validation cohort of BraTS 2020 dataset (125 subjects), as elaborated in Section 3.3, the corresponding feature names were used to extract radiomics features from the discovery dataset. All non-informative features were removed before stability filtering via MAD filtering.

(a) MRMR: We found that superior predictive performance was obtained for an optimal subset of 10 features including age, 2 first-order, and 7 texture features. The statistics of the selected features were: mean OCCC = $0.97 \pm 0.02$ and mean uAUC = $0.52 \pm 0.03$.

(b) RFE-SVM: We found that superior predictive performance was obtained for an optimal subset of 10 features including Age, 1 first-order, and 8 texture features. The statistics of the selected features were: mean OCCC = 0.97 $\pm$ 0.01 and mean uAUC = 0.5 $\pm$ 0.05.

*Important Note:* It must be noted that the validation cohort of BraTS 2020 dataset was only used to identify radiomics features stable to variations in segmentations independently obtained with seven CNNs. Feature names associated with the subset of stable features were used to mine radiomics features from the discovery dataset which were ultimately used for subsequent feature selection and model training. *Hence, no information leakage occurred between model training and inference phases*.

**Table 9.** Predictive performance of radiomics models, quantified with AUROC, for Overall Survival classification task. Each radiomics model corresponds to a specific combination of feature selection method, deep segmentation network, and stability filtering. For inference on the testing dataset (31 subjects), a soft voting method was adopted to unify the outputs of 50 random forest classifiers (with uniform weighting scheme) and generate a single prediction (short-term vs medium-term vs long-term survivor). Bold font indicates best predictive performance (higher AUROC) for each radiomics model.

| Feature Selection Method | Stability filtering status | Number of features | DeepMedicRes 3D CNN | Dong 2D U-Net | Wang 2.5D U-Net | Isensee 3D U-Net | Pereira 3D-2D U-Net | HDC 3D-Net | E1D3 3D U-Net | STAPLE Fusion |
|---|---|---|---|---|---|---|---|---|---|---|
| **MRMR** | True | 10 | **0.69** | 0.66 | **0.72** | 0.68 | 0.70 | 0.69 | 0.69 | 0.71 |
|  | False | 10 | 0.48 | 0.57 | 0.46 | 0.56 | 0.59 | 0.59 | 0.65 | 0.53 |
| **RFE-SVM** | True | 10 | **0.59** | 0.56 | 0.57 | 0.58 | 0.58 | 0.59 | 0.57 | **0.61** |
|  | False | 10 | 0.42 | 0.43 | 0.41 | 0.49 | 0.52 | 0.51 | 0.54 | 0.47 |

*3.5.2. Performance evaluation.* Table 9 summarizes the predictive performance of two feature selection methods, with and without *stability filtering*, and across eight segmentation schemes (seven CNNs and one STAPLE-fusion method) using AUROC. Statistical comparisons of AUROCs were made using the fast DeLong method and robustness of radiomics models, across eight segmentation schemes, was quantified with Relative Standard Deviation (RSD) of AUROCs.

*Without Stability Filtering:* The average predictive performance of the two feature selection methods, across eight segmentation schemes, were as follows: MRMR – AUROC 0.55 $\pm$ 0.06 and RFE-SVM – AUROC 0.47 $\pm$ 0.05. The stability of the models was 11.2% with MRMR and 10.4% with RFE-SVM as measured with RSD, across the eight segmentation methods.

(a) MRMR: With MRMR feature selection method, radiomics model trained with predicted segmentations from $E_1D_3$ 3D U-Net yielded superior overall predictive performance i.e., AUROC 0.65. More specifically, it reported the best predictive performance for medium-term survivors (AUROC 0.54) and long-term survivors (AUROC 0.64) while maintaining good predictive performance for short-term survivors (AUROC 0.64). Radiomics model trained with predicted segmentations from Isensee 3D U-Net yielded best predictive performance for short-term survivors (AUROC 0.70).

(b) RFE-SVM: With RFE-SVM feature selection method, radiomics model trained with predicted segmentations from $E_1D_3$ 3D U-Net yielded superior overall predictive performance i.e., AUROC 0.54. Radiomics model trained with predicted segmentations from Isensee 3D U-Net yielded best predictive performance for long-term survivors (AUROC 0.74). Radiomics model trained with predicted segmentations from HDC 3D-Net yielded best predictive performance for medium-term survivors (AUROC 0.62). Radiomics model trained with predicted segmentations from Pereira 3D-2D U-Net yielded best predictive performance for short-term survivors (AUROC 0.64).

*With Stability Filtering:* The average predictive performance of the two feature selection methods, across eight segmentation schemes, were as follows: MRMR – AUROC 0.69 $\pm$ 0.02 and RFE-SVM –

AUROC 0.58 ± 0.02. The stability of the models was 2.65% with MRMR and 2.67% with RFE-SVM as measured with RSD, across the eight segmentation methods.

(a) MRMR: With MRMR feature selection method, radiomics model trained with predicted segmentations from Wang 2.5D U-Net yielded superior overall predictive performance i.e., AUROC 0.72. More specifically, it reported strong predictive performance for short-term survivors (AUROC 0.76), medium-term survivors (AUROC 0.32) and long-term survivors (AUROC 0.79).

(a) RFE-SVM: With RFE-SVM feature selection method, radiomics model trained with predicted segmentations from STAPLE Fusion yielded superior overall predictive performance i.e., AUROC 0.61. More specifically, it reported strong predictive performance for short-term survivors (AUROC 0.65), medium-term survivors (AUROC 0.67) and long-term survivors (AUROC 0.61).

Stability filtering, followed by MRMR guided feature selection, improved predictive performance for short-term and long-term survivors across all segmentation networks. However, predictive performance for medium-term survivors decreased with stability filtering across all segmentation networks. Stability filtering, followed by RFE-SVM guided feature selection, improved predictive performance only for short-term survivors across all segmentation networks. For medium-term survivors, predictive performance decreased across all six out of eight segmentation networks except STAPLE Fusion where it increased from (0.52 to 0.67), with stability filtering. For long-term survivors, predictive performance decreased across six out of eight segmentation networks, with stability filtering. Stability filtering with MRMR guided feature selection, as opposed to its RFE-SVM counterpart, reported improved predictive performance across all segmentation schemes. More specifically, statistically significant improvements in predictive performance was obtained with HDC 3D-Net and STAPLE-fusion method for medium-term survivors ($p < 0.05$).

## 5. Discussion

Several (mini)processes along a medical imaging pipeline induce variability in radiomics features which ultimately compromises predictive performance on downstream tasks. In the context of brain tumor characterization, we focused on two key questions which, to the best of our knowledge, have not been explored so far: (a) stability of radiomics features to variability in multiregional segmentation masks obtained with fully-automatic deep segmentation methods and (b) subsequent impact on predictive performance on downstream tasks – such as IDH prediction and Overall Survival (OS) classification. We further constrained our study to limited computational resources setting which are found in underprivileged, remote, and/or resource-starved clinical sites in developing countries.

We employed seven state-of-the-art CNNs which can be trained with limited computational resources (GPU with 12 GB RAM) and have demonstrated superior segmentation performance on Brain Tumor Segmentation challenge (BraTS 2015 – 2021). The chosen segmentation architectures include DeepMedicRes, Dong 2D U-Net, Wang 2.5D CNN, Isensee 3D U-Net, Pereira 3D-2D U-Net, HDC 3D-Net, and $E_1D_3$ 3D U-Net. We also employed STAPLE-fusion method to fuse the segmentation labels independently obtained with the seven CNNs. We excluded deep segmentation networks with large memory footprints (GPU with ≥ 40 GB RAM). Empirical analysis revealed that Isensee 3D U-Net was ranked significantly higher in comparison to the remaining six CNNs (FRS = 1, $p < 0.001$). The STAPLE fusion method was ranked second (FRS = 2) and was not significantly lower than Isensee 3D U-Net ($p = 0.49$). Amongst the three tumor subregions, enhancing core (ENC) was the hardest to segment due to poor contrast and fragmented (physiologic) structure (Shaheen et al., 2022). This is evident from lower DSC (67 − 78%) and higher HD-95 (22.9 − 45.5 mm) across eight segmentation schemes (including STAPLE-fusion method).

Closest to our empirical study, on the stability of radiomics features, is the work of (Suter et al., 2020). (Suter et al., 2020) quantified the robustness of radiomics features (using ICC) against 125 perturbations (including voxel size, axial spacing, k-space subsampling, additive Gaussian noise, bin width for gray values, and simulated inter-rater manual segmentation). The exploratory dataset included 3D mpMRI scans, acquired at a single center with homogeneous acquisition parameters, of 19 subjects diagnosed with GBM. A total of 8,327 radiomics features were extracted from tumor subregions. (Suter et al., 2020) found that shape features were the most robust (ICC $\geq$ 0.97), followed by first-order features and texture features. Our study deviates from (Suter et al., 2020) as follows: (i) Our exploratory dataset comprised of multi-center, multi-vendor, and multi-parametric MRI scans from 125 subjects. (ii) The 125 subjects include LGGs and HGGs[f]. (iii) A total of 11,129 radiomics features were extracted from tumor subregions. (iv) We studied stability of radiomics features to (realistic) variability in multiregional segmentation maps independently obtained with seven fully-automatic state-of-the-art CNNs. (v) We used the overall concordance correlation coefficient (OCCC) which quantified the agreement of a radiomics feature, across multiple independent segmentation schemes, on a scale of [0, 1]. Radiomics features with OCCC $\geq$ 0.95 were marked as highly stable. Our study revealed that highly stable radiomics features were: (1) predominantly texture features (79.1%), (2) mainly extracted from WT region (96.1%), and (3) largely representing T1Gd (35.9%) and T1 (28%) sequences. The average stability, in terms of OCCC, for each feature category was as follows: $0.87 \pm 0.12$ for WT, $0.76 \pm 0.13$ for TC, $0.72 \pm 0.13$ for ENC, and $0.72 \pm 0.11$ for Shape. *Shape features and radiomics features extracted from the ENC tumor subregion had the lowest average stability*. We reached the same conclusions if, instead of OCCC, intraclass correlation coefficient (ICC) was used to quantify the stability of radiomics features (see Supplementary).

We evaluated two feature selection approaches, MRMR and RFE-SVM, to extract a compact subset of discriminatory features for downstream prediction tasks – IDH prediction and OS classification. Superior predictive performance was obtained with an optimal subset of ten features. Applying MRMR or RFE-SVM to the original set of radiomics features (11,129) yielded an optimal subset of ten features which were moderately stable: OCCC $\approx$ 0.77 in IDH prediction task and OCCC $\approx$ 0.66 in OS classification task. This is because the optimization function in MRMR and RFE-SVM does not include any regularization term which penalizes selection of *unstable* or *moderately stable* features. The optimal subset included radiomics features from WT, TC, and ENC subregions. *Stability filtering* was performed in advance to reduce the original set of radiomics features (11,129) to a set of highly stable features (820, OCCC $\geq$ 0.95). Shape features and radiomics features from ENC subregion were filtered out. Feature selection with MRMR or RFE-SVM yielded an optimal subset of ten highly stable features (OCCC $\approx$ 0.97). In IDH prediction task, discriminatory radiomics features came from WT and in OS classification task, eight-of-the-nine radiomics features represented WT. Age was only predictive in OS classification task which has been reported before (Puybareau et al., 2018; Sun et al., 2019; Baid et al., 2020; Ammari et al., 2021; Cepeda et al., 2021; Sacli-Bilmez et al., 2023).

For IDH prediction task, we found that RFE-SVM had significantly more efficacy than MRMR feature selection method. This finding has been confirmed in several studies (Zhang et al., 2018; Ren et al., 2019; Bhandari et al., 2021). Stability filtering, followed by RFE-SVM guided feature selection, significantly improved predictive performance across all segmentation networks ($p < 0.05$). Stability filtering minimized non-physiological variability in predictive models as indicated by an order-of-magnitude decrease in RSD (2.28% without stability filtering vs 0.64% with stability filtering). The non-physiological variability is attributed to variability in segmentation maps obtained with fully-automatic CNNs. Pereira 3D-2D U-Net provided an instructive example: Pereira 3D-2D U-Net had the lowest performance score on the segmentation task (FRS = 8). Without stability filtering, radiomics model, learned with discriminatory features from RFE-SVM, had an AUC = 0.77. Stability filtering significantly increased the AUC = 0.944 ($p < 0.005$), bringing the predictive performance on-par with other radiomics models[g].

---

[f] The proportion of LGGs and HGGs in the 125 subjects is not shared publicly.

For Overall Survival (OS) classification task, we found that MRMR is more effective than RFE-SVM feature selection method. This finding has also been confirmed in several studies (Suter et al., 2018, 2020; Deng et al., 2022). Stability filtering, followed by MRMR guided feature selection, improved predictive performance across all segmentation networks. Stability filtering minimized non-physiological variability in predictive models as indicated by an order-of-magnitude decrease in RSD (11.2% without stability filtering vs 2.65% with stability filtering). The non-physiological variability is attributed to variability in segmentation maps obtained with fully-automatic CNNs. Wang 2.5D U-Net provided an instructive example: Wang 2.5D U-Net had the third lowest performance score on the segmentation task (FRS = 6). Without stability filtering, radiomics model, learned with discriminatory features from MRMR, had an AUC = 0.46. Stability filtering increased the AUC = 0.72, bringing the predictive performance on-par with other radiomics model.

Another important empirical finding is associated with the top-ranked segmentation network: Isensee 3D U-Net (FRS = 1). Stability filtering, followed by RFE-SVM feature selection, significantly improved predictive performance for IDH: AUC = 0.83 to AUC = 0.935 ($p < 0.05$). Stability filtering, followed by MRMR feature selection, improved predictive performance for OS classification task: AUC = 0.56 to AUC = 0.68. These results showcase the importance of *suboptimal* deep segmentation networks which can be exploited as *auxiliary networks* for subsequent identification of radiomics features stable to variability in automatically generated multiregional segmentation maps.

Our study had several limitations. The testing dataset for the overall survival classification is quite imbalanced with only three subjects belonging to the medium-term survivors class. However, this does not impact the conclusions we draw about robustness (stability) of radiomics features across fully-automatic deep segmentation networks. Robustness of radiomics features was evaluated on a diverse cohort of 125 subjects. Our exploration was only limited to the BraTS dataset. BraTS is the benchmark dataset for segmentation of tumor subregions on 3D mpMRI scans. BraTS dataset is well-researched and continues to expand with novel downstream tasks. This allows comparative analysis with published literature (Bakas et al., 2018, 2023; Crimi and Bakas, 2021) . For the segmentation task, we focused on seven fully-automatic deep segmentation networks, based on CNNs, which can be trained with limited computational budgets. We excluded deep segmentation networks with large memory footprints; which includes transformer-based or diffusion-guided segmentation networks (GPU with $\geq$ 40 GB RAM). Computational constraints were enforced to simulate limited (computational) infrastructure in underprivileged, remote, and/or resource-starved clinical sites in developing countries.

## 6. Conclusion

In the context of brain tumor characterization, we focused on two key questions which, to the best of our knowledge, have not been explored so far: (a) stability of radiomics features to variability in multiregional segmentation masks obtained with fully-automatic deep segmentation methods and (b) subsequent impact on predictive performance on downstream tasks – such as IDH prediction and Overall Survival (OS) classification. Our study revealed that highly stable radiomics features were: (1) predominantly texture features, (2) mainly extracted from WT region, and (3) largely representing T1Gd and T1 sequences. Shape features and radiomics features extracted from the ENC tumor subregion had the lowest average stability. Stability filtering minimized non-physiological variability in predictive models as indicated by an order-of-magnitude decrease in the relative standard deviation (RSD) of AUCs. The non-physiological variability is attributed to variability in multiregional segmentation maps obtained with fully-automatic CNNs. Stability filtering significantly improved predictive performance on the two downstream tasks, i.e., IDH prediction and Overall Survival classification, substantiating the inevitability of learning novel radiomics and radiogenomics models with stable discriminatory features. The study (implicitly) showcased the importance of suboptimal deep segmentation networks which can be exploited as auxiliary networks for subsequent identification of radiomics features stable to variability in automatically generated multiregional segmentation maps.

---

[g] Other radiomics models refers to radiomics models learned with discriminatory features extracted from 3D mpMRI scans using multiregional segmentation maps obtained with fully-automatic deep segmentation networks.

## Acknowledgements

H.M. was supported by a grant from the Higher Education Commission of Pakistan as part of the National Center for Big Data and Cloud Computing and the Clinical and Translational Imaging Lab at LUMS. The authors are grateful to Syed Talha Bukhari whose extensive work on brain tumor segmentation (Bukhari and Mohy-ud-Din, 2021), conducted at LUMS, provided a strong basis for this study.